\documentclass[preprint]{elsarticle}

\usepackage{amssymb}
\usepackage{indentfirst}
\usepackage{array}
\usepackage{lineno,hyperref}
\modulolinenumbers[5]
\usepackage{graphicx}
\graphicspath{{./figures}}
\usepackage{subfigure}
\usepackage{float}
\usepackage{amsmath}
\usepackage{booktabs}
\usepackage{gensymb}
\usepackage{multirow}
\usepackage{algorithm}
\usepackage{algpseudocode}
\usepackage{changepage}
\usepackage{xcolor}
\usepackage{cleveref}
\usepackage{amsfonts}
\usepackage{dsfont}
\usepackage{mathrsfs}
\usepackage{doi}
\usepackage{color}
\usepackage{multicol}
\usepackage{pdflscape}
\usepackage{algorithm}
\usepackage{algpseudocode}
\usepackage{bm}
\usepackage{lscape}

\journal{}

\begin{document}

\begin{frontmatter}





\title{Identifying Key Drivers of Heatwaves: A Novel Spatio-Temporal Framework for Extreme Event Detection}

\author[uah]{J. P\'erez-Aracil\corref{j}}
\author[uah]{C. Pel\'aez-Rodr\'iguez} 
\author[cmcc]{Ronan McAdam}
\author[cmcc]{Antonello Squintu}
\author[uah]{Cosmin M. Marina}
\author[uah]{Eugenio Lorente-Ramos}
\author[jlu]{Niklas Luther}
\author[bsc]{Verónica Torralba}
\author[cmcc]{Enrico Scoccimarro}
\author[cmcc]{Leone Cavicchia}
\author[polimi]{Matteo Giuliani}
\author[hereon]{Eduardo Zorita}
\author[hereon]{Felicitas Hansen}
\author[csic]{David Barriopedro}
\author[ucm]{Ricardo García-Herrera}
\author[uco]{Pedro A. Gutiérrez}
\author[jlu]{Jürg Luterbacher}
\author[jlu]{Elena Xoplaki}
\author[polimi,cmcc2]{Andrea Castelletti}
\author[uah]{S. Salcedo-Sanz} 

\address[uah]{Department of Signal Processing and Communications, Universidad de Alcal\'a, Alcal\'a de Henares, 28805, Spain.}
\address[cmcc]{CMCC Foundation - Euro-Mediterranean Center on Climate Change, Italy}
\address[polimi]{Department of Electronics, Information, and Bioengineering, Politecnico di Milano, Milano, Italy}
\address[hereon]{Helmholtz-Zentrum Hereon, Hamburg, Germany}
\address[jlu]{1Justus Liebig University Giessen, ZEU}
\address[csic]{Instituto de Geociencias (IGEO), Consejo Superior de Investigaciones Científicas–Universidad Complutense de Madrid, Madrid, Spain.}
\address[ucm]{Departamento de Física de la Tierra y Astrofísica, Facultad de Ciencias Físicas, UCM, Madrid, Spain.}
\address[uco]{Department Informatics and Numerical Analysis, Universidad de Córdoba, Córdoba, Spain.}
\address[bsc]{Barcelona Supercomputing Center (BSC), Barcelona, Spain}
\address[cmcc2]{RFF-CMCC European Institute on Economics and the Environment, Euro-Mediterranean Center on Climate Change, Milano, Italy}

\cortext[j]{Corresponding author: Jorge P\'erez-Aracil {Department of Signal Processing and Communications, Universidad de Alcal\'a, Alcal\'a de Henares, 28805, Spain. e-mail address:~jorge.perezaracil@uah.es}}

\begin{abstract}
Heatwaves (HWs) are extreme atmospheric events that produce significant societal and environmental impacts. Predicting these extreme events remains challenging, as their complex interactions with large-scale atmospheric and climatic variables are difficult to capture with traditional statistical and dynamical models. This work presents a general method for driver identification in extreme climate events. A novel framework (STCO-FS) is proposed to identify key immediate (short-term) HW drivers by combining clustering algorithms with an ensemble evolutionary algorithm. The framework analyzes spatio-temporal data, reduces dimensionality by grouping similar geographical nodes for each variable, and develops driver selection in spatial and temporal domains, identifying the best time lags between predictive variables and HW occurrences. The proposed method has been applied to analyze HWs in the Adda river basin in Italy. The approach effectively identifies significant variables influencing HWs in this region. This research can potentially enhance our understanding of HW drivers and predictability.
\end{abstract}


\begin{highlights}

\item We propose a framework which identifies key short-term heatwave drivers.

\item The proposed method combines dimensionality reduction and evolutionary algorithms.

\item Drivers include climate variables with time lags up to 6 months the event.

\item Dimensionality reduction in the spatial domain is considered.

\item Results at Adda river basin identify key drivers and influential time frames.

\end{highlights}


\end{frontmatter}


\section{Introduction}\label{sec:sample1}

The occurrence of heatwaves (HWs), characterized by prolonged periods of abnormally high temperatures exceeding typical local conditions, has become a pressing concern in recent years due to their severe societal and environmental impacts \cite{perkins2015review,easterling2000observed}.
Since 1950, extensive regions worldwide have witnessed numerous prolonged and intense HWs, resulting in significant consequences for human mortality, regional economies, and natural ecosystems \cite{meehl2004more,zittis2022climate,kuglitsch2010heat, della2007doubled, garcia2010review}. In agriculture, heat stress on crops can significantly reduce yields, leading to food insecurity. In addition, increased demand for electricity for cooling during HWs substantially strains power grids. The escalation in the frequency of HWs has been documented in various parts of the globe in recent years and is at least partly attributed to the temperature increases driven by anthropogenic warming \cite{russo2014magnitude,russo2015top}.

Numerous studies \cite{barriopedro2011hot,sillmann2013climate,stott2011single} have consistently highlighted that the ongoing increase in global surface temperatures will lead to significant alterations in the frequency and intensity of HWs across Europe by the end of this century. This trend is not confined to Europe; globally, there is also a growing prevalence of heat extremes, with projections indicating that these events will continue to increase in the coming decades
\cite{battisti2009historical,coumou2013historic,fischer2013robust}. Regional differences can be encountered in HW projections. Hence, this leads to diverse drivers and climate forcings on regional scales. The identification of these drivers plays a key role in understanding regional variations and in developing effective mitigation and adaptation strategies, as different regions may experience distinct climate impacts due to a combination of local factors and global climate forces. Moreover, understanding these drivers is crucial for improving forecasts on sub-seasonal scales, allowing for more accurate predictions of HWs and other extreme events.

When tackling the challenge of HW detection or prediction, it is necessary to understand the mechanisms responsible for these extreme events. Although the underlying processes remain not entirely understood \cite{perkins2015review}, an increasing number of studies have delved into these mechanisms and physical drivers that contribute to the formation and prediction of HWs \cite{domeisen2023prediction,barriopedro2023heat}. HWs are the product of intricate interactions between large- and small-scale processes that operate across diverse temporal scales. These events are highly influenced by atmospheric circulation, often regarded as a fast-acting driver, as well as anomalous conditions in slowly changing climate components, which can serve as proximate factors (e.g., land surface) or remote factors (e.g., upper ocean temperature, or sea ice) affecting HWs occurrence \cite{sillmann2017understanding,hoskins2015persistent,miralles2019land}. In the extratropics, atmospheric circulation patterns that influence HWs include quasi-stationary synoptic-scale high-pressure systems (anticyclones)  \cite{brunner2018dependence,sousa2018european}, whose predictability at a seasonal scale is low due to the influence of the chaotic variability of the atmosphere \cite{prodhomme2021seasonal}.  Finally, long-term trends in frequency, duration, and intensity of HWs are primarily driven by anthropogenic forcings, including global factors such as greenhouse gas concentrations and regional factors like land-use/land-cover changes and aerosol emissions \cite{seneviratne2021weather}. However, these are out of the scope of this paper.

In close relation to the previous discussion, and considering the vast volume of available spatial and temporal data, employing data-driven methodologies becomes indispensable for uncovering potential HW drivers.
A limited body of literature addresses this subject using ML and feature selection and dimensionality-reduction approaches. Some works \cite{asadollah2021prediction,buschow2023explaining,loughran2017understanding} employed Principal Components Analysis (PCA) to reduce and optimize the number of highly correlated variables, using them as inputs in some ML algorithms. In \cite{wehrli2019identifying}, authors aimed to identify the role of the individual drivers for five HWs in the recent decade through factorial experiments, which force the model toward observations for one or several key components at a time, allowing to identify how much of the observed temperature anomaly of each event can be attributed to each driver. Other feature selection approaches have been used for different weather problems in searching for optimal input variables. In \cite{tao2022integration}, an extreme gradient boosting feature selection algorithm was applied with ML models in a problem of short-term relative humidity prediction. In \cite{hagen2021identifying}, a nested loop of roughly pruned random forests was used for identifying significant drivers of daily streamflow from large-scale atmospheric circulation in Norway. In \cite{chaqdid2023extreme}, a clustering method was applied to divide Morocco into regions that are spatially consistent in terms of extreme precipitation and to identify its drivers by analyzing atmospheric circulation anomalies during the occurrence of regional events. In \cite{orimoloye2022drought}, ML regression-based algorithms were used to identify the drivers of drought dynamics in the Free State Province. \cite{dalal2024drivers} shows the influence of different drivers to understand the causal mechanism of HWs over South-West India. For that purpose, climate model simulations and long-term observational data were proposed.


This study proposes a general framework for HW driver identification, which can be applied to other extreme events in the context of detection and event short-term prediction. The framework is illustrated here to detect HWs in a European location. Specifically, the framework proposed in this work follows a two-phase methodology to obtain robust HW driver identification. In the first phase, a clustering algorithm is applied to variables identified as potential drivers, extracted from the ERA5 reanalysis dataset \cite{ERA5}, and presented as time series. This clustering step reduces the dimensionality of the spatial domain by grouping nodes with similar time series patterns. In the second phase, a wrapper feature selection approach based on a multi-method ensemble evolutionary algorithm (PCRO-SL) \cite{perez2023new} is employed to identify the most skilful drivers and periods for HW forecast over short-term (days to weeks) and seasonal horizons. The optimization algorithm’s fitness function performs a driver selection by evaluating the performance of an ML model for HW classification based on a subset of clustered drivers. 

The proposed framework is applied to the agricultural districts in the Adda river basin, located downstream of Lake Como, in the Lombardy region, Northern Italy. These districts are part of the Po Valley, one of the most productive European agricultural areas, which provides one-third of the national agricultural production \cite{eurostat_irrigation}. Understanding the crop risks associated with extreme temperatures is becoming increasingly crucial to planning effective climate change adaptation strategies.



The manuscript is organized as follows. First, a description of the data, including potential drivers and target variables used for developing the experiments, is provided in Section \ref{sec_Data}. Then, the spatio-temporal feature selection methodology is presented and detailed in Section \ref{sec_Methodology}. Subsequently, the experimental work and the results obtained are further described in Section \ref{sec_results}. Finally, in Section \ref{sec_conclusions}, there is a discussion on the potential uses of the framework in wider-scale driver detection and on implications for forecasting.


\section{Data description} \label{sec_Data}

This section provides a detailed description of the data used for the experiments and the construction of the target. First, regarding HW definition, this issue has been widely discussed in the literature \cite{barriopedro2023heat}. This work follows the widely-used HW definition given in \cite{russo2015top} based on cumulative normalized daily maximum temperature (TX) exceedances. Next, we will provide details on the potential drivers considered and the target variables considered.

\subsection{Potential drivers and target} \label{section_drivers}

The variables considered as potential drivers to perform the HW prediction may be categorized into three groups: 
1) meteorological variables, local or remote, 2) climate indices, 3) other variables. 

The first group consists of atmospheric, ocean and other variables which influence climate on various timescales: mean sea level pressure (MSLP), outgoing longwave radiation (OLR), total precipitation (TP), height of the 500 hPa geopotential (Z500), 2m temperature (T2M), as well as sea surface temperature (SST), sea ice concentration (SIC) and volumetric soil moisture in the upper 7cm (SM). 

The second group, climate indices, are included because long-term indices are linked to large-scale atmospheric patterns that influence temperature over extended periods \cite{kenyon2008influence}. In addition, long-term indices help distinguish between climate change variability and natural variability. However, the role of large-scale drivers and teleconnections in the Adda river basin, as for much of Europe, is still unclear \citep{giuliani2019detecting}. 
First, the NINO3.4 index (area-averaged SST anomaly in the region 5$^o$S-5$^o$N, 120$^o$W-170$^o$W) is used to represent the El Nino Southern Oscillation (ENSO), which is strongly linked with the occurrence of extreme heat in northern continents \cite{luo2020summer,luo2019amplifying,martija2021enso}. The Indian Ocean Dipole (IOD), whose association with HWs has been investigated in recent years \cite{reddy2021interactive,dalal2024drivers}, is calculated as the difference of area-averaged SST anomaly between the western tropical Indian Ocean (50$^o$E–70$^o$E, 10$^o$N–10$^o$S) and the southeastern tropical Indian Ocean (90$^o$E–110$^o$E, O$^o$N-10$^o$S). Lastly, the North Atlantic Oscillation (NAO), whose impact on European heatwaves has been previously studied in \cite{li2020collaborative,mukherjee2020compound,kueh20202018},  is derived from the first principal component of Z500 in the North Atlantic domain.

The third group covers miscellaneous variables such as mean atmospheric $CO_2$ levels \cite{lemordant2016modification} and the specific calendar day of the year (DOY) \cite{li2023regional}.


Table \ref{tab_predictors} describes the geographical domain considered for each meteorological variable. Some have been studied in two domains to account for their varying influence in various geographical scales. Also, land variables available over the local region under study are considered as independent potential drivers (MSLP, OLR, SM, T2M, TP and Z500).


\begin{table}[!ht]
     \caption{Predictive variables considered at each node from the ERA-5 reanalysis dataset. The specific coordinates corresponding to the geographical limits: Europe: [30-70N, 16W-44E], Arctic: [48-90N, 180W-180E] and North Atlantic: [0-66N, 90W-40E].}
     \label{tab_predictors}
     \centering
     \begin{tabular}{ c  c  c c} 
      \textbf{\#} & \textbf{Variable} & \textbf{Domain} \\
      \toprule \toprule
        1 & Mean Sea Level Pressure (MLSP) & Global, Europe\\
        2 & Soil Moisture (SM) & Europe & \\
        3 & Sea Ice Cover (SIC) & Arctic &\\
        4 & Sea Surface Temperature (SST) & Global, North Atlantic &\\
        5 & Height of the 500 hPa Geopotential (Z500) & Global, Europe\\
        9 & Total Precipitation (TP) & Europe \\
        10 & Outgoing Longwave Radiation (OLR) & Global, North Atlantic\\
        11 & 2m Temperature (T2M) & Europe\\
              \toprule \toprule
    \end{tabular}
 \end{table}


Regarding the target variable, we have selected the agricultural districts in the Adda river basin (including Lake Como) in the north of Italy (centered around 46º N, 9º E). For this location, a daily time series of binary HW occurrence index over 1950-2022, for the warmest months of the year (May, June, July and August) using the HW definition given in \cite{russo2015top}.

\subsection{Data extraction and preparation }

The detection of HWs presented in this paper is performed based on physical variables data extracted from ERA5 reanalysis \cite{ERA5}. ERA5 provides hourly information on a broad set of variables, such as temperature, pressure, precipitation, and snowfall, with a resolution of 0.25 degrees in both longitude and latitude. Daily average values were considered in this case, with a horizontal resolution of 0.5 degrees. The 72-year ERA5 database, based on data from 1950 to 2022, has been considered for both target and predictive variables. These data have been divided into a training period from 1950 to 2010 and a test period from 2010 to 2022. For the training split, the positive cases (HW occurrence) represent 5.1\% of the cases, while for the test split, the positive cases represent 15.1 \% of the cases. 

The climatology is computed for each calendar day using the 1981-2010 period, and a rolling average of 30 days is applied to smooth this annual cycle. The local seasonal cycle is removed for each candidate driver to provide a time series of anomalies.




\section{Spatio-Temporal Cluster-Optimized Feature Selection (STCO-FS)}
\label{sec_Methodology}
\begin{figure}[!ht]

 \makebox[\textwidth][c]{\includegraphics[width=0.7\textwidth]{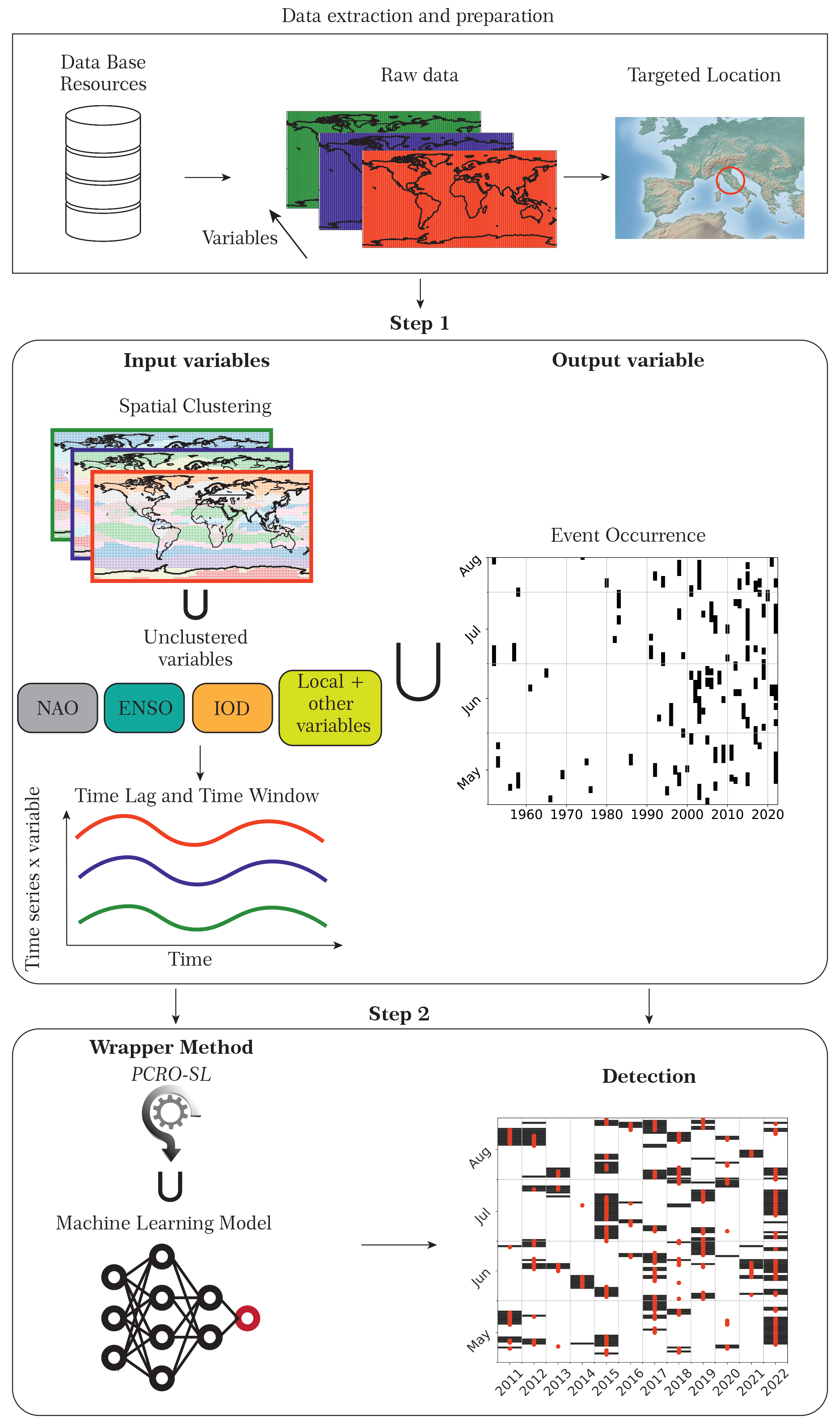}}
    \caption{Scheme of the proposed feature selection framework.}
\label{figure_methodology}
\end{figure}

This section presents the proposed framework to identify optimal HW drivers in spatial and temporal domains. Figure \ref{figure_methodology} illustrates the methodology flow, where it is worthwhile to distinguish between the data treatment of potential predictor variables and the target variable. The possible drivers have been defined in Section \ref{section_drivers}. 
The proposed framework has two steps (plus a preprocessing stage), as shown in Figure \ref{figure_methodology}. The first step consists of clustering the drivers to reduce the spatial dimension. The area-weighted spatial average time series of clusters are then merged with the local variables and climate indices. In the second step, a wrapper feature selection method is applied using an evolutionary optimization algorithm and a ML method for selecting the optimal time frames of each potential driver. 

Once the input and target variables have been processed, they are used to feed supervised ML classification algorithms, which conduct the detection of the HW occurrence over the Adda river basin in the considered period. The framework steps are described in the subsequent sections.

\subsection{Dimensionality Reduction through Clustering}

The clusters represent geographic areas in which the temporal variability is most similar. Previous studies dictate the use of different domains for certain variables. SM and precipitation are considered local-scale influences on European HWs \cite{stefanon2012heatwave}. European temperatures are indirectly linked to Arctic SIC \citep{zhang2018skillful} \citep{sun2022influence} through impacts on atmospheric circulation, but we assume no link to Antarctic SIC. North Atlantic SSTs affect the European climate, particularly in winter \citep{gastineau2015influence}, but there is some evidence to suggest a lagged influence on summer heatwaves \citep{duchez2016drivers,bischof2023role}. Meanwhile, global modes of climate variability, which influence distant continents via teleconnections, are represented by global SST patterns. For an explicit consideration of atmospheric dynamics and teleconnections, z500 and MSLP are used locally and globally. Extreme heat is often linked to regional scale circulation anomalies \citep{sousa2018european}, which can be excited and/or amplified remotely via atmospheric teleconnections \citep{barriopedro2023heat}.

The classic K-means clustering algorithm has been employed to reduce the spatial dimensionality of the predictive data. Initially introduced by MacQueen \cite{macqueen1967some}, K-means has become one of the most widely used and extensively studied clustering algorithms. The key input parameter in the K-means approach is the number of clusters, denoted by $k$. The algorithm then partitions the data into $k$ clusters by following defined steps. It is important to note that the proposed methodology allows other possible clustering algorithms (different from K-means) to cluster the input data. 

The proposed method allows the inclusion of any other clustering method that better fits the problem under study. It is important to note that, at this stage of the process, the critical point is to reduce the dimensionality of the problem, particularly in the spatial domain.

    
    

This study implements the clustering algorithm for the 12 predictors outlined in Table \ref{tab_predictors}. Next, the area-weighted spatial average for each cluster is computed. Consequently, a time series is generated for each cluster under consideration. A value of $k=5$ clusters has been considered for each predictor field, giving a total of 60 clusters involved in the prediction $(5 \text{ clusters} \times 12 \text{ variables} = 60 \text{ potential drivers})$. Clustering was applied considering the training data (from 1950 to 2010) of daily time series anomalies relative to 1981-2010. For each variable, five clusters are obtained, displayed in Figure \ref{fig_clusters}. For T2M, values over land are only considered. For SIC, areas historically free of sea ice have been masked out.


Although the number of clusters is arbitrary, it is essential to note that this work aims to present the methodology but not its best configuration, which may depend on the location of the study and the different variables considered. 



The proposed method allows the introduction of unclustered variables as potential drivers of the problem. Thus, those variables that do not need to be spatially grouped can be selected, and their time series are included in the analysis. This study considers 11 additional features included as potential drivers, including the climate indexes, the local meteorological variables (MSLP, OLR, SM, T2M, TP and Z500 taken over the Adda river basin) and the other variables defined in Section \ref{section_drivers}. Thus, these results in a database of predictor variables comprising 71 variables arranged as a time series.

\subsection{Candidate Selection: the optimization problem}

\begin{figure}[!ht]
  \begin{center}
        \includegraphics[width=1\textwidth]{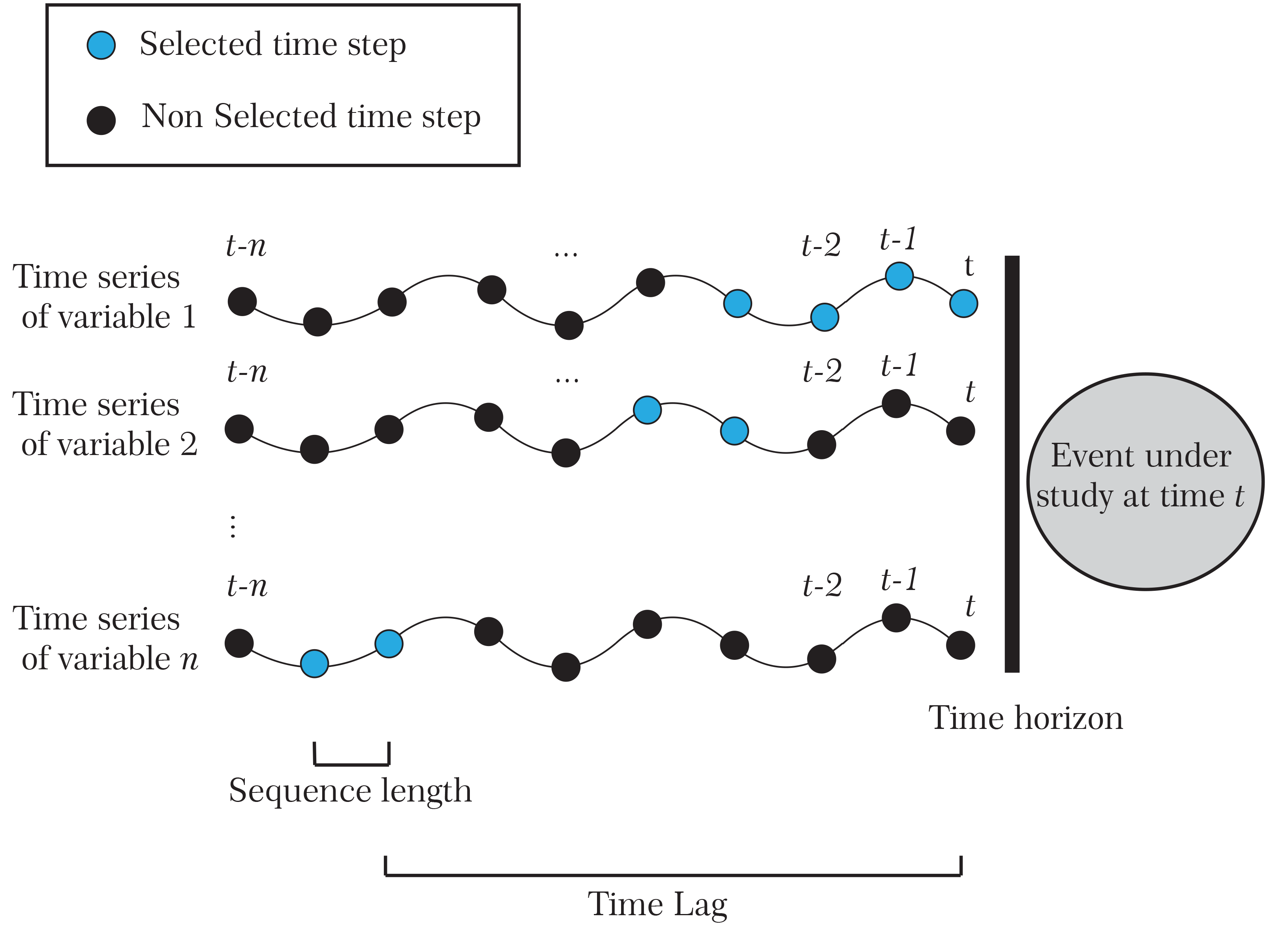}
    \caption{Illustration of time series: time lag, forecast horizon and sequence length parameters. }
    \label{figure_time_lag}
  \end{center}
\end{figure}

After reducing the dimensionality of the problem in the spatial domain, our focus shifts to the temporal dimension. Here, the objective is to identify periods and lags exhibiting the highest predictive skill for each potential driver. 
First, the prediction time-horizon is defined, determining how far in advance predictor data should be considered. This work sets the time horizon to zero since it is configured as a detection problem. Based on the prediction horizon, the time lag and sequence length values are searched for each potential driver under study using evolutionary computation. They represent the lead time and the window length considered for each variable, which enables us to distinguish between short-term (low time lag) and long-term (high time lag) predictors. An illustration is shown in Figure \ref{figure_time_lag}. In this work, the maximum time lag is set as 180 days, and the maximum window length is set as 60 days. Therefore, the evolutionary search could account for a lead time of up to 8 months for each of the variables.

The process of determining the time lag and the sequence length has been conducted using a robust and well-established optimization algorithm: the Probabilistic Coral Reef Optimization Algorithm with Substrate Layers (PCRO-SL) \cite{salcedo2017review,perez2023new}. It is a low-level ensemble for optimization \cite{wu2019ensemble}, based on evolutionary computation. It was first proposed as an advanced version of the original CRO algorithm \cite{salcedo2014coral}, which was an evolutionary-type meta-heuristic, proposed as a class of hybrid between Evolutionary Algorithms \cite{del2019bio} and Simulated Annealing \cite{kirkpatrick1983optimization}. 

The PCRO-SL algorithm further evolves the CRO approach towards a multi-method ensemble. It generally proceeds as the original CRO, but with a significant difference: Instead of having a single way of evolving, it considers several {\em substrate layers} of the approximately same size in the reef. Each substrate, in turn, represents a particular evolution strategy or searching procedure. Thus, the PCRO-SL is a multi-method ensemble algorithm \cite{wu2019ensemble}, where several searching strategies are carried out within a single population.

Different combinations of well-known meta-heuristics may be implemented. In this case, we considered regular combinations of previously defined meta-heuristics. Specifically, we have defined and applied the following substrates in the PCRO-SL: Harmony Search (HS), Multipoint crossover(MPx), XOR operation (XOR) and BLX-$\alpha$ crossover (BLX).

The optimization problem formulation for selecting the optimal time-domain features for each driver is structured as follows. Each solution generated by the PCRO-SL consists of an array containing three key variables for each candidate driver: time lag, sequence length, and a binary indicator determining whether the driver is included or discarded. Including this binary variable encourages solutions that prioritize minimal information, helping reduce the impact of noise. The time lag is constrained within $[h, 180]$ days, where h is the time horizon, while the sequence length is limited to $[1, 60]$ days. In this initial study, we focus on a prediction time horizon of $h=0$ days, aiming to identify drivers informing on HW detection (nowcasting) up to time scales longer than six months. The variation in sequence length allows us to investigate drivers that maintain significant influence over different periods within the time domain.

The achieved F1 score dictates the optimal lag selection for each variable. The F1 score is the harmonic mean of the measures: precision (ratio of correctly predicted positives to all predicted positives) and recall (ratio between the correctly predicted positives to all observed positives). The F1-score is widely used to assess the quality of binary predictions. In our case, the F1-score achieved by each candidate driver served as the fitness function for selecting optimal clustered and unclustered variables, with their corresponding time lags and sequence length, and other variables. To accomplish this, a deterministic and fast training classifier, namely the popular Logistic Regressor (LR) \cite{kleinbaum2002logistic}, was used (other ML algorithms were tested). When the optimization algorithm provides a potential solution (comprising three values per driver), the chosen time lags are concatenated into a tabular format, and LR training is performed. This process was conducted using a Cross-Validation (CV) approach: the entire training data is divided into five validation folds, and the average error encountered for these folds has been used as the fitness function of the optimization algorithm.

\subsection{Machine Learning classifiers} \label{sec_ML_methods}

Although a fast-training ML algorithm such as LR is used during the optimization process, a variety of more sophisticated models are subsequently implemented to evaluate the optimal solution the PCRO-SL algorithm provides. These models include: Light Gradient Boosting Machine (LGBM) \cite{ke2017lightgbm}, Support Vector Classifier (SVC) \cite{Scholkopf2000}, Decision Trees (DTs) \cite{loh2011classification}, Random Forest (RF) \cite{Breiman2001}, Gaussian Naive Bayes (GNB) \cite{Zhang2004}, K-Nearest Neighbours (KNN) \cite{Mucherino2009}, Adaptive Boosting (AB) \cite{Freund1996}, Multi-Layer Perceptron (MLP) \cite{Gardner1998}, Gradient Boost (GB) \cite{friedman2001greedy} and Extreme Learning Machine (ELM) \cite{Huang2006}.

These methods are implemented in Python using the following libraries: \texttt{sklearn}, \texttt{skelm} and \texttt{lightgbm}. The hyperparameters of these classifiers are determined using a random hyperparameter search with the values considered in Table \ref{MethodsParameters}. A CV with five folds was performed. 

\begin{table}[!ht]
     \caption{Parameters of the experimental setup. The hyperparameters of each model are described for the models used in this work: Light Gradient Boosting Machine (LGBM), Support Vector Classifier (SVC), Decision Trees (DTs), Random Forest (RF), Gaussian Naive Bayes (GNB), K-Nearest Neighbours (KNN), Adaptive Boosting (AB), Multi-Layer Perceptron (MLP), Gradient Boost (GB),  Extreme Learning Machine (ELM).}
     \footnotesize
     \label{MethodsParameters}
     \centering
     \begin{tabular}{ c l l |  l l  } 
      
          \toprule \toprule

        \multirow {13}{3cm}{\centering ML Methods} & \textbf{LGBM} & ~ & \textbf{SVC} & ~ \\ \cline{2-5}
       
       & num leaves & 20-200 & C & 0.1-1000  \\

       & n estimators & 50-500 & Gamma & 0.001-1  \\

       & ~ & ~ & Kernel & \textit{rbf}  \\
       
       & \textbf{DT} & ~ & \textbf{RF} & ~ \\ \cline{2-5}
       
       & max depth & 1-50 & n estimators & 100-600  \\
       
       & min samples leaf & 1-50 & bootstraps & True/False  \\
       
       & \textbf{GNB} & ~ & \textbf{KNN} & ~ \\ \cline{2-5}
       
       & var smoothing & -9-0 & n neighbors & 3-30  \\
       
       & \textbf{AB} & ~ & \textbf{ELM} & ~ \\ \cline{2-5}
       
       & n estimators & 50-200 & n neurons &  10-500 \\
       
       & learning rate & 0.001-10 & ~ & ~  \\

        & \textbf{GB} & ~ & \textbf{MLP} & ~ \\ \cline{2-5}
       
       & n estimators & 50-300 & n layers &  1-4 \\
       
       & learning rate & 0.01-0.2 & n neurons & 32-512  \\

       & max depth & 1-9 & activation & \textit{relu}  \\

       & ~ & ~ & solver & \textit{adam}  \\

    & ~ & ~ & alpha & 0.0001-0.01  \\

    & ~ & ~ & batch size & 16-64  \\

    & ~ & ~ & learning rate & 0.0001-0.01  \\

    & ~ & ~ & max iters & 200-600  \\
       
       \bottomrule \bottomrule   
 
     \end{tabular}
 \end{table}

\section{Experimental work and results} \label{sec_results}

This section describes the experimental work, the results obtained, and the corresponding discussion. First, Section \ref{results_optimization} details the drivers the optimization algorithm selects. Second, Section \ref{results_ML} shows the results provided by the ML models in the detection task. In all experiments, the training period spans from 1951 to 2009, while the test period covers 2010 to 2020, comprising 18\% of the total data. The training dataset identifies optimal drivers and model training, whereas the test dataset is reserved exclusively for evaluating model performance.

\begin{figure}
    \centering
    \begin{adjustwidth}{0cm}{0cm}
        \includegraphics[width=1\textwidth]{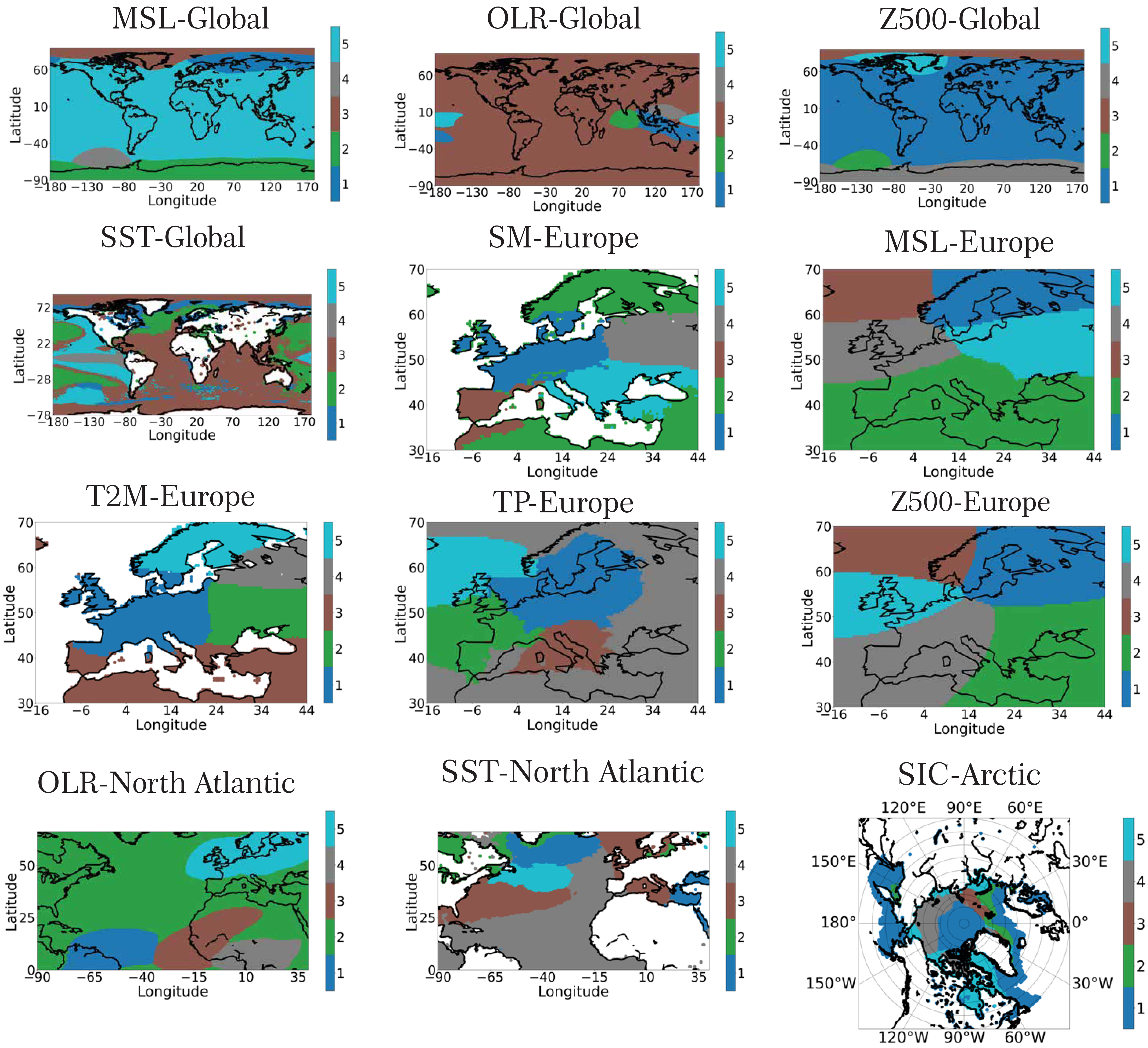}
    \caption{Clusters provided by the K-Means+ algorithm ($K=5$) for the first group of variables: meteorological variables.}
    \label{fig_clusters}
        \end{adjustwidth}
\end{figure}

\subsection{The selected variables} \label{results_optimization}

The optimization algorithm is initialized once the potential predictors are made up, including clusters, unclustered variables, and the local and climate indices. The clustered variables are shown in Figure \ref{fig_clusters}.

The solution provided by the proposed method is a vector of length $3 \times \text{number of variables}$. It represents the time lag, sequence length and a binary variable with the time steps for which the variable is selected. In this case, the forecast horizon has been set to 0 since the goal is to validate the methodology for unveiling potential drivers before attempting a more complex forecasting problem.

\begin{figure}[!ht]
\begin{adjustwidth}{-5cm}{-5cm}
    \centering
    \includegraphics[clip,width=1.5\textwidth]{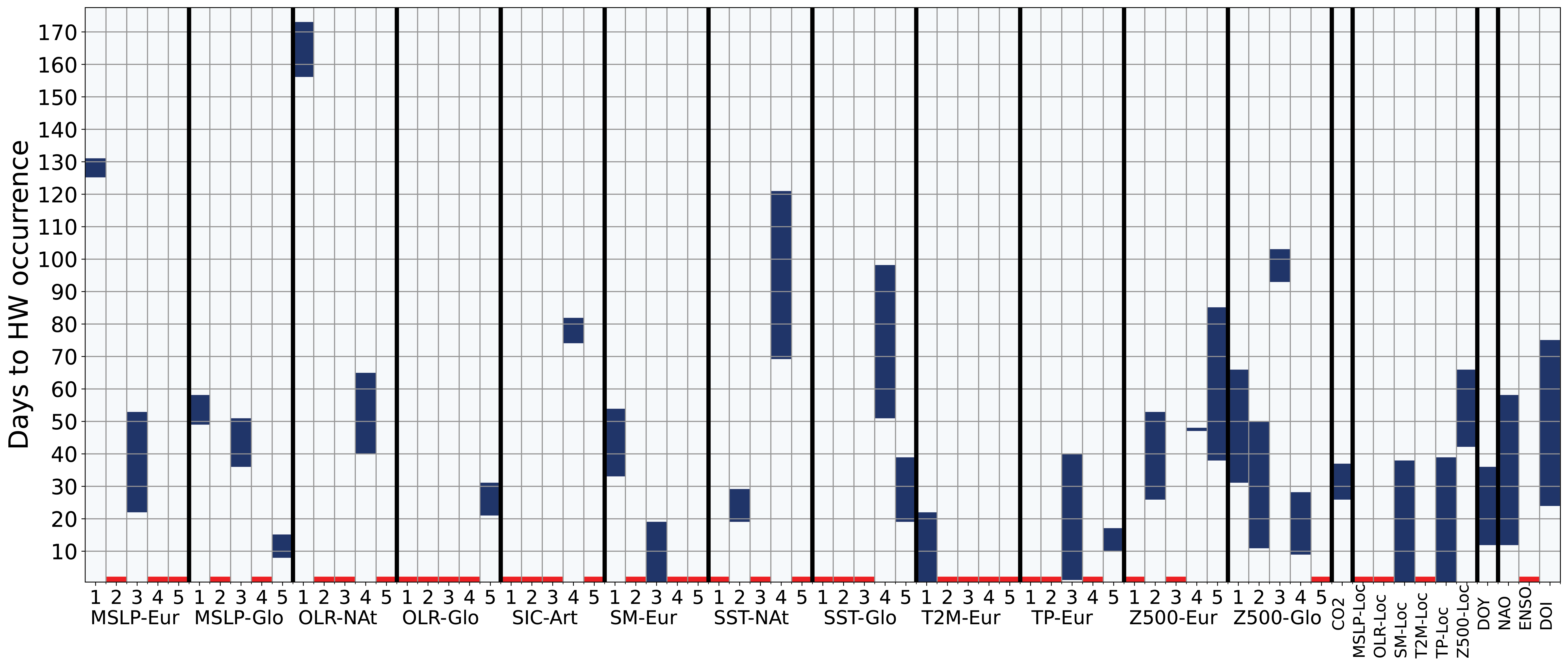}
    \end{adjustwidth}
    \caption{Example of the individual solution provided by the optimization algorithm. Red boxes represent variables that have not been selected in this particular solution.}
    \label{fig_bestsolution}
\end{figure}

The PCRO-SL optimization algorithm is executed in ten independent runs to prevent false positives caused by the inherent randomness in this optimization problem. Figure \ref{fig_bestsolution} shows a potential solution the algorithm provides, corresponding to the best solution of all the runs, in terms of CV error. Here, in the x-axis, the 71 potential drivers are listed. The y-axis shows the temporal scale (in days relative to the HW occurrence). This plot highlights in blue the time steps the optimization algorithm selects for each predictor in the case of this specific example. The red square means that the algorithm has discarded the specific potential driver. 

Figure \ref{fig_bestsolution} helps the user to interpret, visually and intuitively, the different variables that are being chosen, as well as distinguish between variables that have a sort term influence (in this case, T2M in the European domain over the cluster 1, SM in the European domain, cluster 3, and TP in the European domain over the cluster 3 (see Figure \ref{fig_clusters}), and the T2M and Z500 at the local node of Lake Como); and others variable with a delayed impact that influence with a delay of up to 170 days (e.g. OLR over the North Atlantic domain in cluster 1).

\begin{figure}[!ht]
  \begin{center}
         \includegraphics[clip,width=1\textwidth]{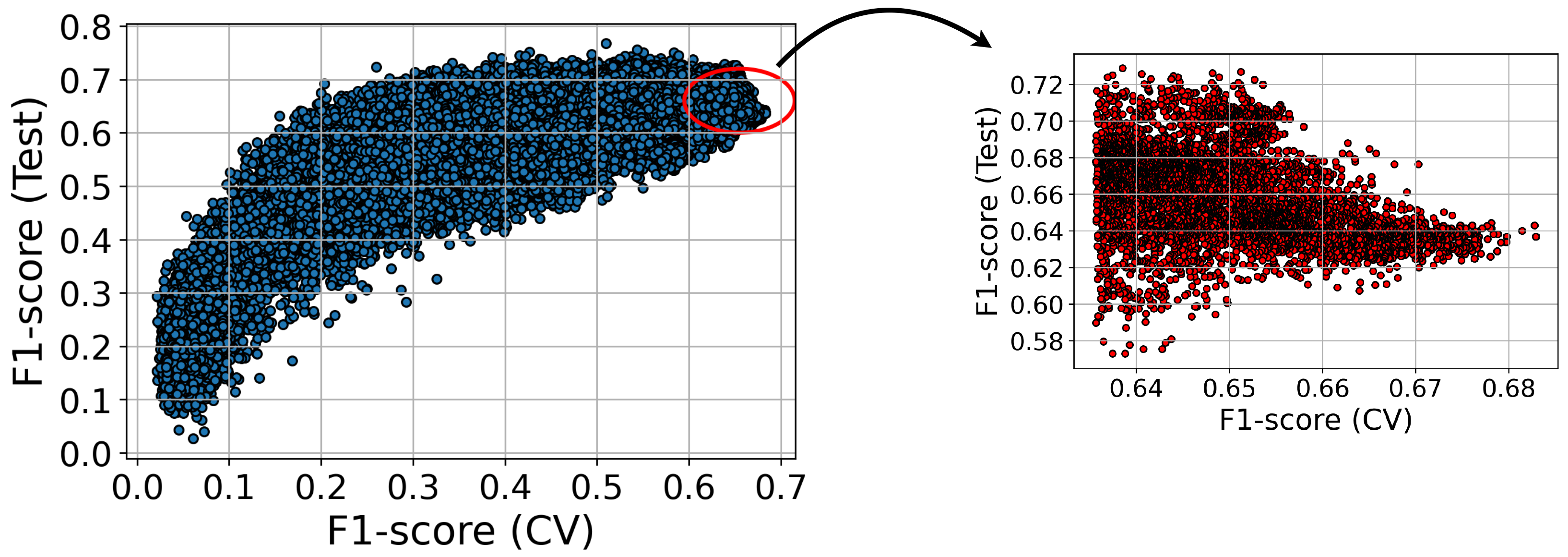}
    \caption{Test vs train (CV) performance of all the potential solutions tried by the optimization algorithm. The solutions represented in red are the ones selected for an in-depth study.}
    \label{fig_scatter}
  \end{center}
\end{figure}

The question that can arise when examining individual solutions is whether or not all the selected time steps provide predictability to the problem or whether they are noise that has been added to the truly significant variables. To analyse this aspect, we evaluated all potential solutions generated by each independent run of the optimization algorithm. A total of 150,000 potential solutions are generated (15,000 evaluations of the fitness function $\times$ 10 independent runs). Each solution represents a combination of predictor variables, tested on the train (by 5-fold CV) and test data. The metrics of each combination of drivers are plotted in Figure \ref{fig_scatter}. Out of these 150,000 possible solutions, the best 10\% (in terms of CV error) have been selected to analyze further the predictor variables that provide significance for the prediction problem under evaluation. These selected combinations are represented with red points in the scatter plot of  Figure \ref{fig_scatter}.

\begin{landscape}
\begin{figure}[!ht]
    \centering
    \includegraphics[clip,width=1.7\textwidth]{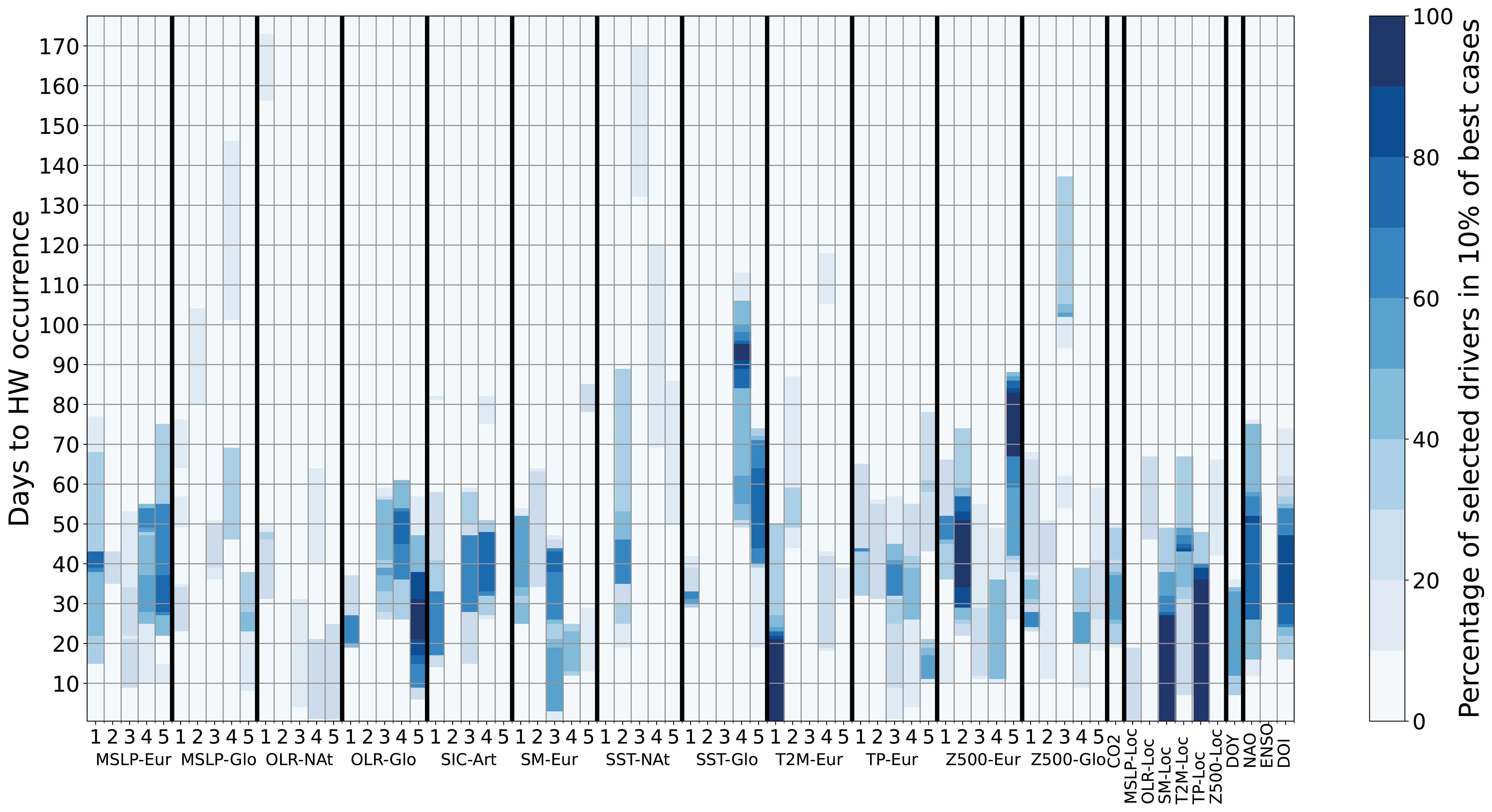}
    \caption{Heatmap representing the best 10\% of solutions the optimization algorithm provides. For a specific time step, darker colours mean that most proposed solutions select the variable, while light colours denote that it is barely selected.}
    \label{fig_heatmap_bests}
\end{figure}
\end{landscape}

The best solutions are then analyzed in the frequency map reported in Figure \ref{fig_heatmap_bests}. The darker the colour, the more frequently that variable has been selected in that time lag. Low intensity means that the corresponding time step has been barely chosen among the best solutions and is therefore considered noise. 

It becomes evident that certain variables selected by some solutions merely introduce noise, obscuring the truly significant variables. Generally, the most frequently selected features are found on the short-term and sub-seasonal time scales (i.e. $<$45 days). Three of the most frequently chosen drivers all have short lag times ($<$20 days): T2M-Eur-1 (the regional cluster in which the Adda basin is found), and local values of SM and TP. These predictors are considered important over 20 days before the event rather than the day, indicating the importance of persisting conditions. The selection of nearby short-term drivers is not surprising, particularly given the previously identified roles of soil moisture and precipitation in summer temperatures \citep{stefanon2012heatwave,ardilouze2019investigating}. On the sub-seasonal scale, OLR-Glo-5 (western-central Pacific; 20-30 days), Z500-Eur-2 (eastern Mediterranean; 30-50 days), NAO and IOD (both over 20-55 days) are the most frequently selected predictors. On the seasonal scale, Z500-Eur-5 (North Atlantic; 70-85 days) and SST-Glo-4 (Tropical Pacific; 90-100 days) are the most selected. While the scope of this study is not to perform a process-based study on how detected features physically impact HW occurrence, it is shown here that the selection of some features is at least partially supported by evidence. For other variables, this framework may act as a first step in understanding which processes need to be studied further. 

Regional-scale atmospheric circulation is considered key, judging by the frequent selection of Z500-Eur 2 (eastern Mediterranean; 30-50 days) and Z500-Eur-5 (British Isles; 70-85 days). However, features such as blocks and ridges are known to determine the occurrence and intensity of summer HWs in the days before an event \citep{sousa2018european}, instead of the subseasonal-to-seasonal (S2S) timescales detected here. Meanwhile, the NAO index, representing circulation over the North Atlantic with impacts on weather across the continent, is also detected as an important predictor on the S2S timescale (20-55 days). Persistence of NAO (specifically, of the positive phase), and of the blocking with which it interacts, has been found before severe heatwave events in the region \citep{drouard2019disentangling,kueh20202018}. It is unclear why NAO was also not selected in the 10 days before HWs. Overall, the selected circulation-based features (in z500 and NAO) likely represent precursor wave trains and potentially very persistent blocking \citep{domeisen2023prediction}, but the exact mechanisms require further inspection. This shows how the framework can provide motivation and direction for further analysis of extreme event drivers.


Other selected features include the IOD, 20-55 days prior, another interconnection with influences on European climate and extremes in particular \citep{behera2013origin}. Lastly, the calendar day (DOY) and, to a lesser extent, the global mean atmospheric $CO_2$ concentration are widely selected and therefore considered important for the algorithm to identify HWs, but the selection of a specific lag time for each is considered arbitrary. 

\begin{figure}[!ht]

    \centering
    \includegraphics[clip,width=0.65\textwidth]{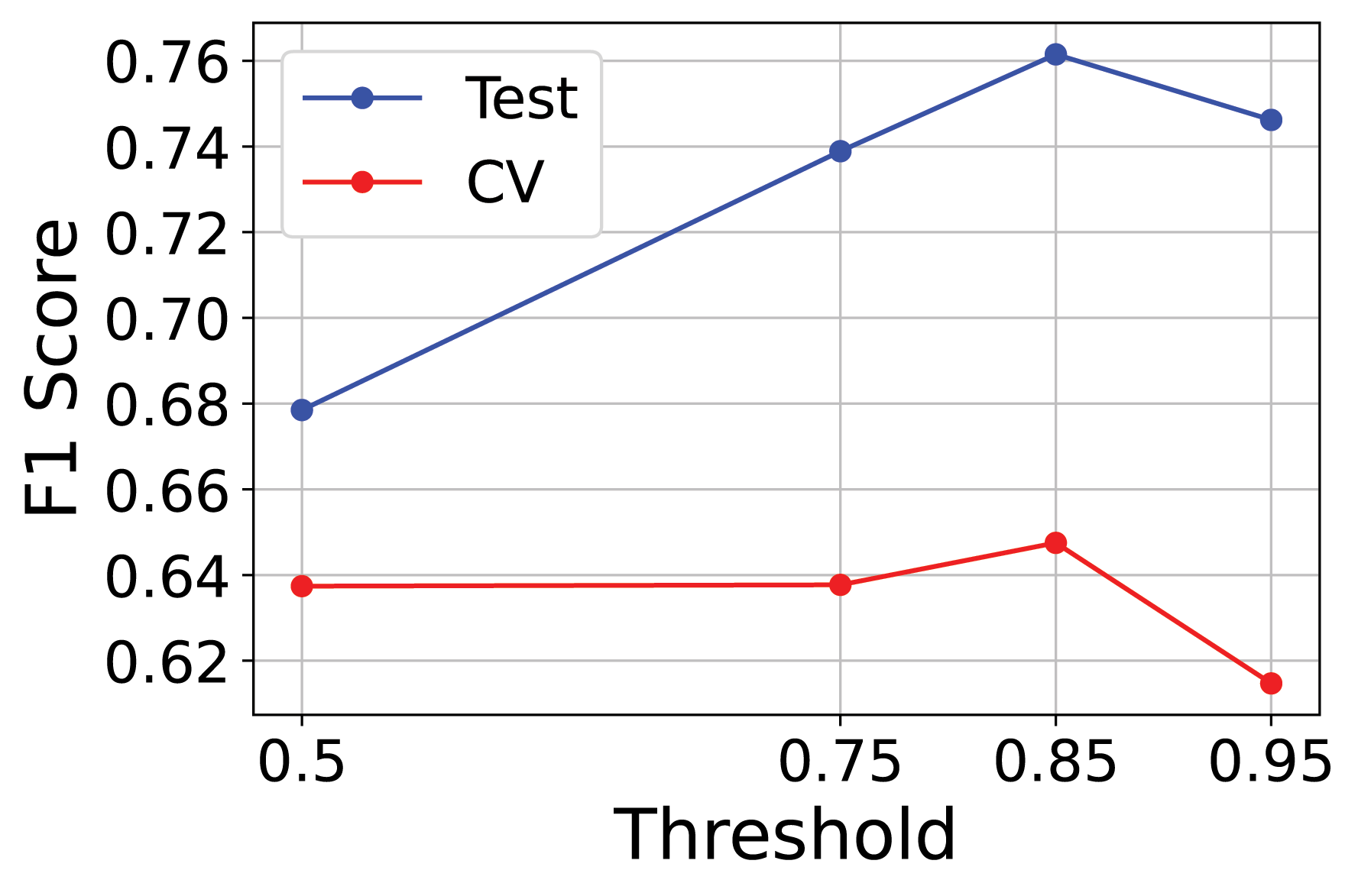}

\caption{Evolution of test F1-scores in HW detection when increasing the agreement threshold (reducing the number of selected drivers).}
    \label{fig_threshold_HW}
\end{figure}

The next step involves determining the variables that contribute robustly as potential drivers. For this purpose, a threshold has been established in the frequency map of the selected variable, such that only those time steps of each variable selected more frequently than the threshold are considered. As the threshold value increases, the number of predictor variables considered decreases, ultimately isolating those consistently present variables in nearly all of the best solutions. The impact of varying the input variables on classifier performance can be observed in Table \ref{table_F1_vs_threshold_HW}, which presents the evolution of the F1-score metric on the test dataset as the threshold increases. The threshold is expressed in percentiles: for instance, a threshold of 0.5 indicates selecting features that appear in at least 50\% of the solutions depicted in the heat map shown in Figure \ref{fig_heatmap_bests}. Figure \ref{fig_threshold_HW} depicts how the performance of the LR classifier improves with the threshold until an upper limit (0.85, i.e. 85 \% of the time is selected), beyond which predictions worsen.

\begin{figure}[!ht]
  \begin{center}
\begin{adjustwidth}{-2cm}{-2cm}
    \subfigure[Threshold: 0.5]{
         \includegraphics[clip,width=0.6\textwidth]{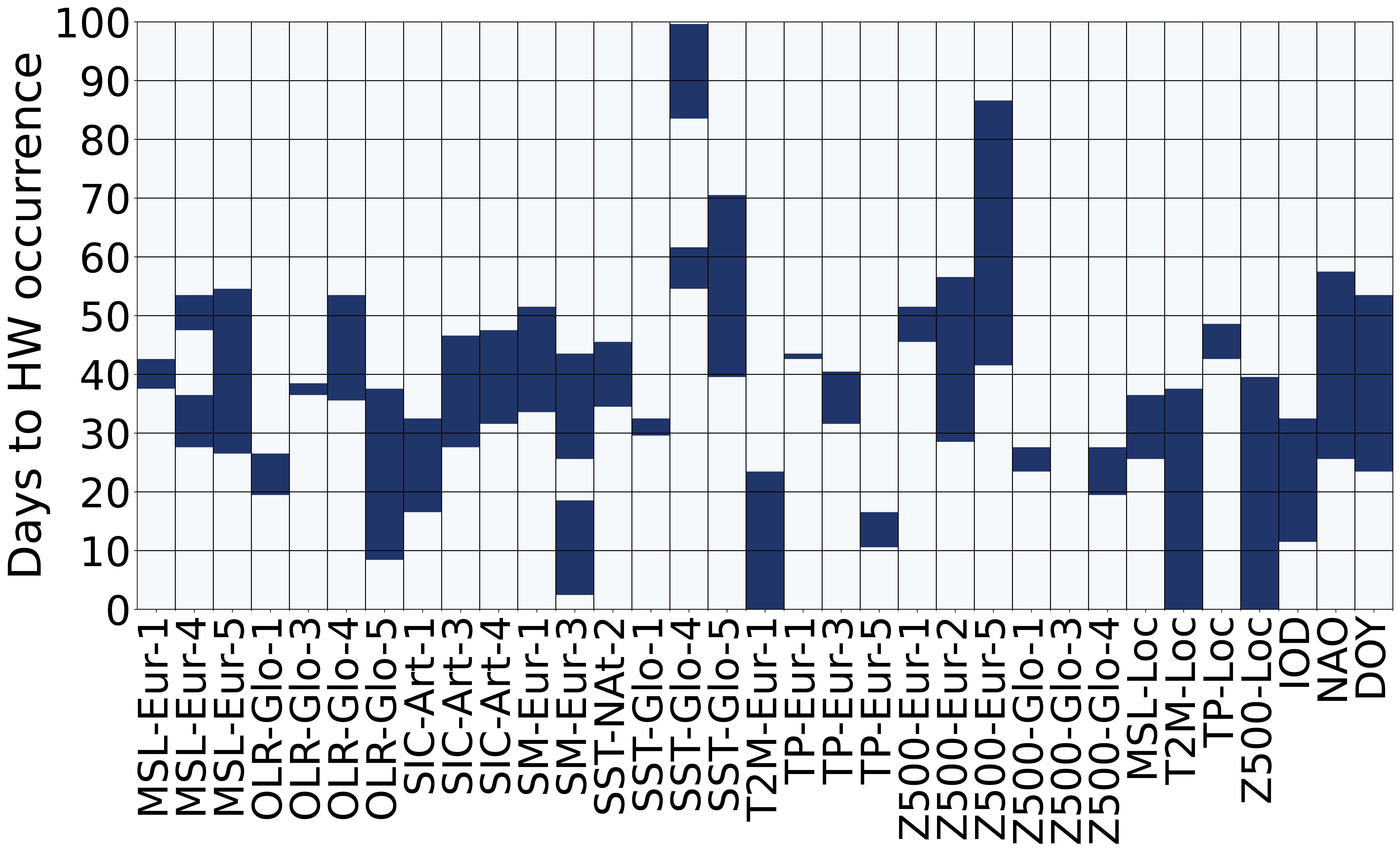}
     }
    \subfigure[Threshold: 0.75]{
         \includegraphics[clip,width=0.6\textwidth]{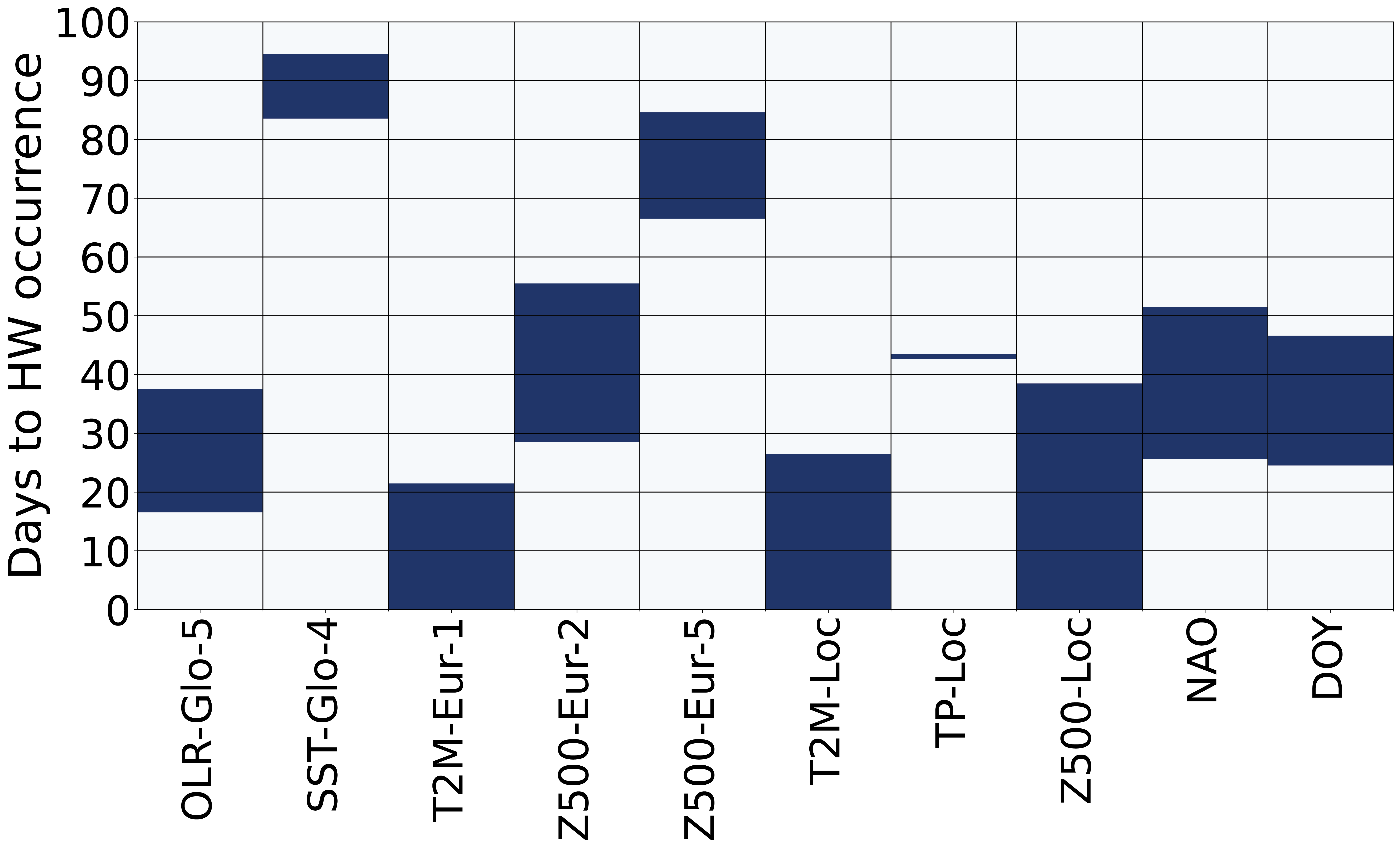}
    }
    \subfigure[Threshold: 0.85]{
         \label{fig_optimum_threshold_HW}\includegraphics[clip,width=0.6\textwidth]{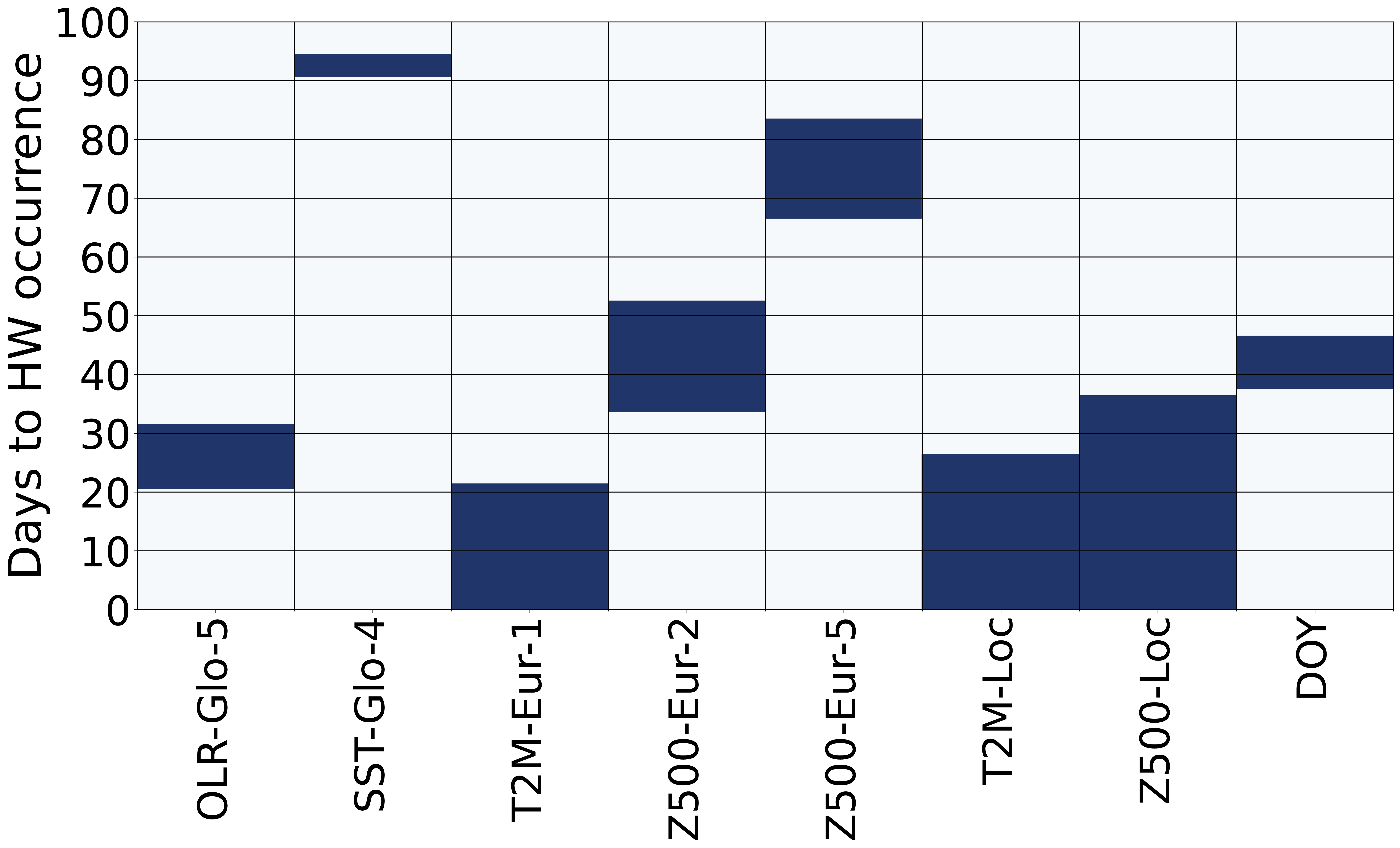}
    }
    \subfigure[Threshold: 0.95]{
         \includegraphics[clip,width=0.6\textwidth]{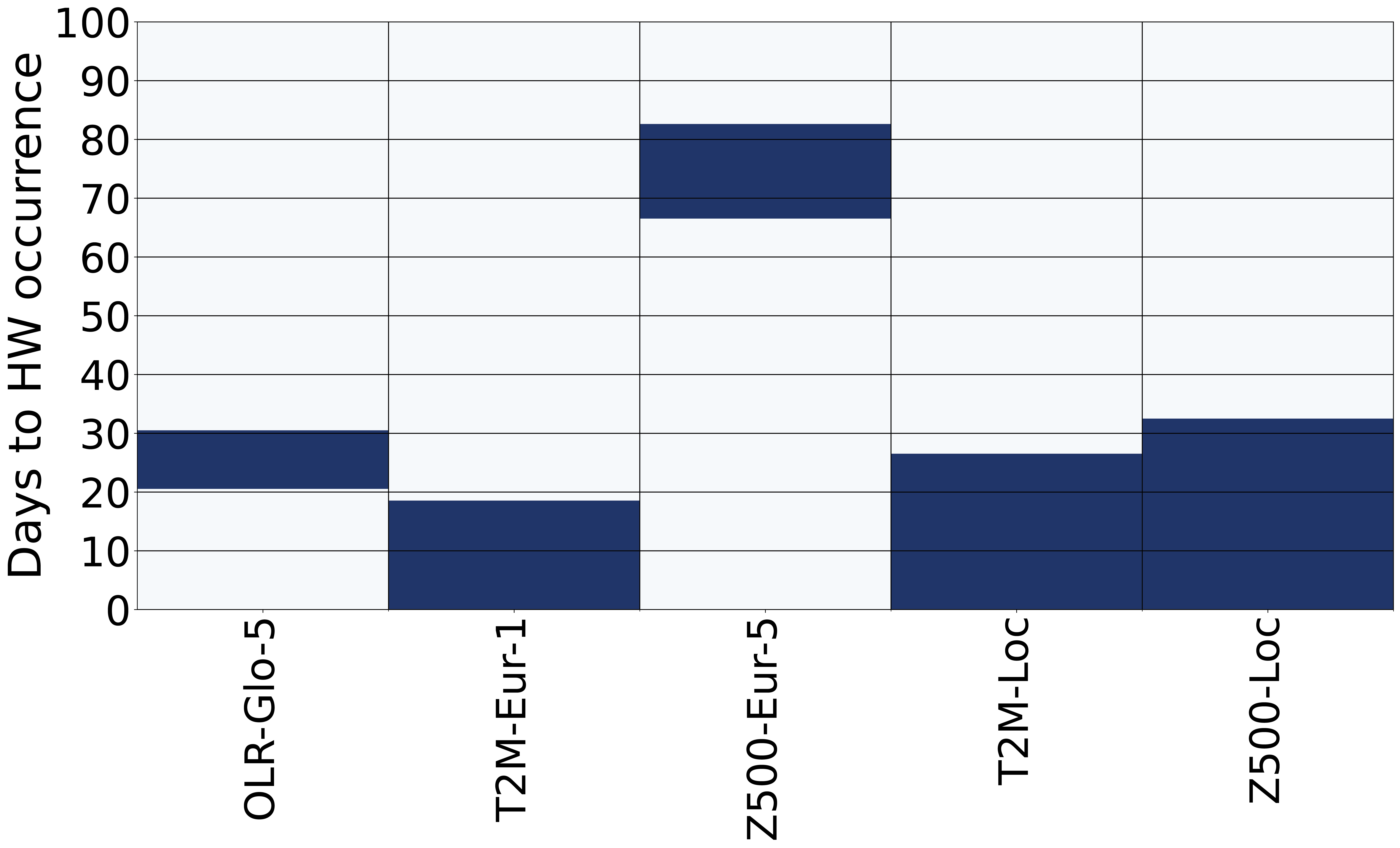}
    }
\end{adjustwidth}
    \caption{Features that are most often chosen by the solutions proposed by the optimization algorithm, in the case of HW index. A threshold of 0.5 represents that these variables appear in at least 50\% of the 15000 best solutions.}
    \label{fig_features_threshold_HW}
  \end{center}
\end{figure}

The group of input variables that are involved in the optimum threshold (0.85, Figure \ref{fig_optimum_threshold_HW}) is the same as previously indicated when analyzing Figure \ref{fig_heatmap_bests}, with some variables concerning the local conditions at short term, and other variables involving broader geographical scales and remote regions with predictive skill on the medium and long ranges (e.g. SSTs).

\begin{table}[!ht]
    \footnotesize
    \caption{Evolution of test F1-scores when increasing the agreement threshold in the selected drivers across the experiments.}
    \label{table_F1_vs_threshold_HW}
    \centering
    \resizebox{0.75\textwidth}{!}{
    \begin{tabular}{cccc} 
        \toprule \toprule
        \textbf{Threshold} & \textbf{Test F1-score} & \textbf{CV F1-score} & \textbf{Nº of features} \\
        \midrule
        0.50& 0.6785 & 0.6374 & 585\\
        0.75& 0.7389 & 0.6377 & 214\\
        0.85& 0.7615 & 0.6475 & 146\\
        0.95& 0.7462 & 0.6147 & 105\\
        \bottomrule \bottomrule
    \end{tabular}}
\end{table}

\subsection{Results for different Machine Learning models} \label{results_ML}

\begin{figure}[!ht]
  \begin{center}
         \includegraphics[clip,width=0.85\textwidth]{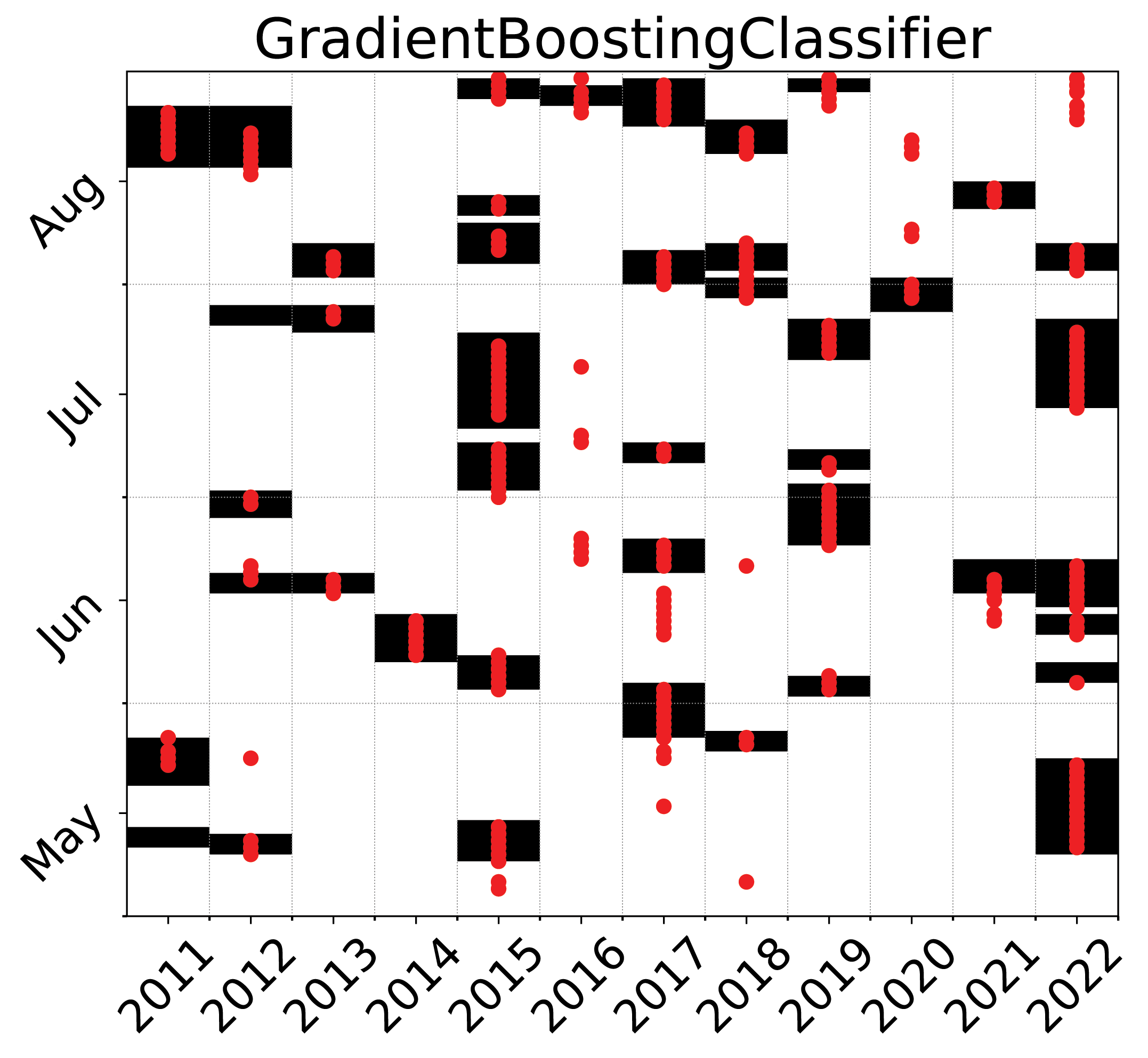}
    \caption{Results from the top-performing ML classifier (GB) over the test period, considering the HW index as the target variable. Black boxes correspond to days with extreme temperature, while red circles denote days predicted by the model as extreme}
    \label{fig_GB_HW}
  \end{center}
\end{figure}

\begin{figure}[!ht]
  \begin{center}
         \includegraphics[clip,width=1\textwidth]{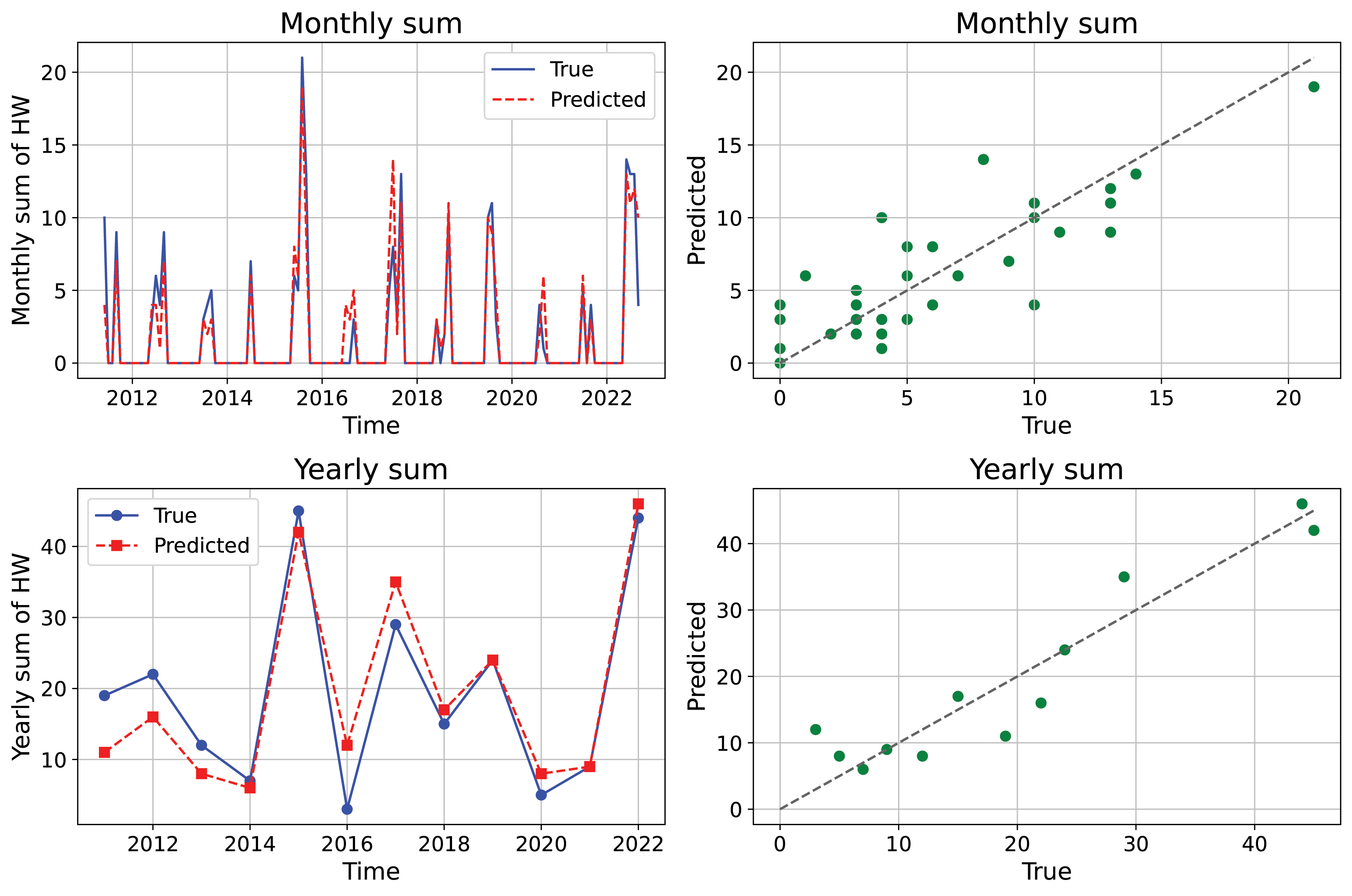}
    \caption{Results provided by the GB model over the test period, considering the HW index as the target variable. Each year, the frequency of actual and predicted HW days is accumulated monthly (top figures) and seasonal (bottom figures). The figures on the left represent the real and predicted time series. The right figures show the scatter plot between the actual vs predicted HW days, featuring a high correlation. }
    \label{fig_forecastplot_GB_HW}
  \end{center}
\end{figure}

Finally, the optimum combination of drivers, corresponding to a threshold equal to 0.85, is used to train a pool of ML classifiers (Section \ref{sec_ML_methods}). The test error metrics for these methods are shown in Table \ref{table_ML_HW}. The predictive results provided by the best-performing model, GB (an F1-score of 0.7906), are further detailed in Figures \ref{fig_GB_HW} and \ref{fig_forecastplot_GB_HW}, which demonstrate the excellent performance of the classification task and the accuracy of the model to detect the HW days over the test domain, with consistent results in all the months and years analyzed.

\begin{table}[!ht]
    \footnotesize
    \caption{Test error metrics for the different ML classification methods assessed, considering the HW index as the target variable. Models are trained with the selected features corresponding to threshold 0.85.}
    \label{table_ML_HW}
    \centering
    \resizebox{0.5\textwidth}{!}{
    \begin{tabular}{cccc} 
        \toprule \toprule
         & \textbf{Recall} & \textbf{Precision} & \textbf{F1-score} \\
        \midrule
        LR & 0.7436 & 0.7803 & 0.7615\\
         LGBM & 0.7094 & 0.8177 & 0.7597\\
        SVC& 0.7051 & 0.7933 & 0.7466\\
        DT& 0.7350 & 0.7818 & 0.7577\\
        RF & 0.6068 & \textbf{0.8658} & 0.7136\\
        GNB& \textbf{0.9658} & 0.3435 & 0.5067\\
        KNN & 0.0983 & 0.7931 & 0.1749\\
        AB& 0.7009 & 0.7923 & 0.7438\\
        MLP & 0.6966 & 0.7376 & 0.7165\\
        GB& 0.7906 & 0.7906 & \textbf{0.7906}\\
        ELM& 0.1752 & 0.7736 & 0.2857\\
        \bottomrule \bottomrule
    \end{tabular}
    }
\end{table}

\section{Conclusions and further research} \label{sec_conclusions}

Understanding the drivers behind the formation of Heatwaves (HWs) is vital to enhance our ability to anticipate, forecast and mitigate the impacts of these extreme events, ultimately reducing the risks to human health, economies, and ecosystems. Researchers can develop more sophisticated models that improve predictive capabilities by recognizing the complex interactions between atmospheric conditions, oceanic patterns, and terrestrial processes. This enhanced understanding also informs the development of effective adaptation and mitigation strategies, ensuring societies are better equipped to handle the increasing frequency and severity of HWs driven by climate change.

This study proposes a comprehensive framework to investigate the interactions between HWs and potential physical drivers across multiple spatio-temporal scales (STCO-FS). The proposed methodology follows a two-phase approach: initially, a clustering algorithm is applied to reduce spatial dimensionality by grouping similar time series data from the ERA5 reanalysis database. Additional variables, such as climate indices and local meteorological factors, are then incorporated into the database. In the second phase, a multi-method ensemble evolutionary algorithm (PCRO-SL) identifies significant periods and clusters relevant to HW occurrences. This approach determines which variables are crucial for HW prediction and the specific time frames in which they are most influential, distinguishing between short-term and long-term drivers. 

The framework has been successfully applied to an agriculturally intensive region in North Italy, demonstrating its ability to detect key HW drivers effectively. A standard definition of HW has been considered. Regarding the potential drivers considered, 8 variables covering different geographic domains, together with three climate indexes (ENSO, NAO and IOD) and local meteorological conditions, have been incorporated into the study. The results indicate strong HW detection capabilities, with nowcasting error metrics of 0.8363. 

For the specific drivers identified in each case, relationships have been established between the occurrence of heat waves and various variables across different spatio-temporal scales. The key selected predictors for the Adda River basin are regional-scale temperature and atmospheric circulation in the 20 days before the event and ENSO, IOD and NAO on sub-seasonal to seasonal timescales. While some of these connections have been suggested in previous work, this study indicates the locations and time frames in which specific mechanisms can be studied and is thus a powerful tool and first step in understanding processes and predictability.

The proposed method offers a series of advantages that are outlined as follows: (1) It enables the discovery of new potential drivers in both temporal and spatial domains simultaneously; (2) By grouping meteorological variables into clusters, the dimensionality of the problem is significantly reduced, and the level of granularity can be adjusted as a parameter of the algorithm; (3) Encoding the problem by establishing lag time and the length of the selected window for each predictor variable reduces the dimensionality in the time domain, thereby simplifying the evolutionary process; (4) The use of a robust evolutionary algorithm (PCRO-SL), along with a fast, efficient, and deterministic classifier, allows the resolution of a complex optimization problem within a short period.
    
Future lines of research will focus on several key areas. Firstly, the framework will be extended to predict HWs with a longer prediction horizon (i.e., S2S) and to tackle predicting the number of HWs on a seasonal scale. Additionally, future research will examine how the selected drivers vary depending on the region under investigation and the data they are applied to (e.g., historical or future climate simulations). This continued research will further enhance the predictive power and applicability of the framework, contributing to more effective HW management and mitigation strategies. Moreover, given the flexibility and modularity of the framework, both predictors and target data can be changed, meaning it can be applied to other extremes with ease. How the clusters are created can also be analysed. Unconnected clusters often arise from variability unrelated to the seasonal cycle. (i.e. not removed when calculating the anomaly). A new way to avoid this is being under study. 

\section*{Code and data availability}
\url{https://github.com/GheodeAI/STCO-FS.git}

\section*{Acknowledgements}
This work has been partially supported by the European Commission, project ``CLImate INTelligence: Extreme events detection, attribution and adaptation design using machine learning, CLINT'' (grant ref.: H2020-LC-CLA-2020-2, 101003876), and by the ``Agencia Estatal de Investigación (España)'', Spanish Ministry of Research and Innovation through NEXO project (grant ref.:  PID2023-150663NB-C21).

 \bibliographystyle{elsarticle-num} 





\end{document}